\documentstyle[11pt]{article}
\let\oldsection=\section
\def\section#1{\oldsection{#1}\indent}
\vspace{6cm}

\title{\LARGE{
   A model calculation of the value of the electromagnetic coupling
   constant at $q^2 = m_{Z}^{2}$ }}

\author{ B.V.Geshkenbein, V.L.Morgunov}

\date{\small{May 30, 1994}}

\begin{document}

\maketitle

\scriptsize{E-mails: geshken@vxitep.itep.msk.su , morgunov@vxdesy.desy.de}

\vspace{3cm}

\abstract{
A QCD model with an infinite number of vector mesons suggested by one
of the authors is used to derive the value of the correction
$\delta\alpha_{hadr}$ for $\alpha(m_{Z}^{2})$ due to the strong interactions.

The result is $\delta\alpha_{hadr} = 0.0275(2)$ ;
thus $\alpha(m^{2}_{Z}) = (128.96(3))^{-1}$. }

\vspace{6cm}


The aim of this work is to obtain the contribution
of strong interactions to the value of $\overline{\alpha} = \alpha(m_{Z}^{2})$.
The more exact knowledge of the value of $\overline{\alpha}$ allows to improve
the predictions for the $t$-quark and Higgs boson masses [1-4].

The quantity $\overline{\alpha}$ is defined by the following formula:
\begin{equation}
 \overline{\alpha} = \frac{\alpha}{1 - \delta \alpha}
\end{equation}
Contributions of $e$, $\mu$ and $\tau$ in (1) are readily calculated
[5] with the result:
\begin{equation}
    \delta \alpha_l = \frac{\alpha}{3 \pi} \left[ \sum_{.}^{.} \ln
      \frac{m_{Z}^{2}}{m_{l}^{2}} - \frac{5}{3} \right] =
      \frac{\alpha}{3 \pi}
      \left[ 22.5 + 11.8 + 6.2 \right]  = 0.0314
\end{equation}
Hadronic contribution in (1) $\delta \alpha$ is given by the
following formula [6]
\begin{equation}
    \delta\alpha_{hadr} = \frac{\alpha m_{Z}^{2}}{3\pi} \; \wp \! \! \!
    \int\limits_{4 m_{\pi}^{2}}^{\infty} \frac{R(s) ds}{(m_{Z}^{2} - s) s}
\end{equation}
The value of  $\delta \alpha_{hadr}$ was obtained in [5,7]
\begin{equation}
    \delta \alpha_{hadr} = 0.0282(9)
\end{equation}
by using the experimental cross-section $e^{+} e^{-} \rightarrow$ hadrons
below $s_0 = (40 GeV)^2$ and the parton model above $s_0$ .
A simple way to understand the value of $\delta \alpha_{hadr}$ can be
found in [8].

We will use a QCD model with an infinite number of vector mesons
[9-11] for theoretical determination of $\delta \alpha_{hadr}$.

Let us write the value $R$ in the form:
\begin{equation}
   R = R_{I=1} + R_{I=0} + R_s + R_c + R_b
\end{equation}
Where $R_{I=1}$ and $R_{I=0}$ is the contribution of $u$ and $d$ quarks
in the state with isotopic spin $I=1$ ($\rho$ - family) and $I=0$
($\omega$ - family)
and $R_s$, $R_c$, $R_b$ are the contributions of $s$ $(\varphi$ - family),
$c$ ($J/\psi$ - family) and $b$ ($\Upsilon$ - family) quarks respectively.

Let us consider for example $J/\psi$ family. In the approximation of
an infinite number of narrow resonances, having masses $M_{k}$
and electronic widths $\Gamma_{k}^{ee}$ , the function $R_{c}(s)$ has the form:
\begin{equation}
    R_{c}(s) = \frac{9 \pi}{\alpha^2} \sum_{k=0}^{\infty} \Gamma_{k}^{ee}
    M_{k} \; \delta (s - M_{k}^{2} )
\end{equation}
Where $\alpha^{-1} = 137.0359895$ [12].

The contribution of $R_c$ in $\delta \alpha_{hadr}$ has the form:
\begin{equation}
 \delta \alpha_{c} = \frac{3 m_{Z}^{2}}{\alpha} \sum_{k=0}^{\infty} f(s_{k})
\end{equation}
Where
\[
f(s_k) = \frac{\Gamma_{k}^{ee} M_{k}}{(m_{Z}^{2} - s_k )s_k} \; \;
, \; \; s_k = M_{k}^{2}
\]

If for $k > 5$ the total widths $\Gamma_{k}$ and masses $M_{k}$
of the vector mesons obey the conditions
\begin{equation}
   M_{k}^{2} - M_{k-1}^{2} \ll M_{k} \Gamma_{k}
      \ll M_{k}^{2}
\end{equation}
then for $k > 5$ the function $R_{c} (s)$ will be described by a smooth curve
and all the formulae of the model [9-11] can be used.

We transform the sum in (7) into an integral by means of the Euler-Maclaurin
formula [13] beginning from $k = k_0$. Choose $k_0 = 4$.
\begin{eqnarray}
  \sum_{k=4}^{\infty} f(s_{k}) = I + \frac{1}{2}f(s_{4}) -
      \frac{1}{12}f^{(1)}(s_{4})+ \frac{1}{720} f^{(3)}(s_{4}) - \nonumber \\
      - \frac{1}{30240}f^{(5)}(s_{4}) + \cdots
\end{eqnarray}
In (9) we have introduced the notations

\[
f^{(l)}(s_{k}) = \frac{\partial^l f(s_k)}{\partial k^l}\mid_{k=4} \; \;
, \; \; I = \int \limits_{s_4}^{\infty} f(s_k) \frac{dk}{ds_k} ds_k
\]

In ref.[10-11] we established a correspondence between the electronic
width of k-th
resonance $\Gamma_{k}^{ee}$ and a derivative of the mass of
k-th resonance $M_{k}$ with respect to the number $k$ of this resonance
[10-11]
\begin{equation}
   \Gamma_{k}^{ee} =  \frac{2 \alpha^2}{9 \pi}  R_{c}^{PT}(s_k)
    \frac{dM_{k}}{dk}
\end{equation}
According to ref.[10-11] the function $R_{c}^{PT}$ includes all gluonic
corrections  in perturbation theory (PT).
The term $ f^{(1)}(s_4)/12$ is approximately equal to
\begin{equation}
    \frac{1}{12} f^{(1)}(s_4) = \frac{1}{24} \left[ f(s_5) - f(s_3) \right]
\end{equation}
This term is small (see table 2). The remaining terms
$f^{(3)}(s_4)/720 - f^{(5)}(s_4)/30240 + \cdots$ may be omitted
due to their smallness. The quantity of $\delta \alpha_{c}$ practically
does not change if $k_0 = 1,2,3$. We can not estimate of the term
$f^{(1)}(s_{k_{0}})/12$ if $k_0 = 0$ or $k_0 = 5$ .

Let us write the integral in (9) in the form (analytical part +
numerical part):
%
\[
I = \frac{\alpha^2}{9\pi} \;
\wp \! \! \int\limits_{s_4}^{\infty}
\frac{R_{c}^{PT}(s) ds}{(m_{Z}^{2}-s)s} =
\]
\[
=  \frac{\alpha^2}{9\pi}
\left\{ R_{c}^{PT}(m_{Z}^{2}) \;
\wp \! \! \int\limits_{s_4}^{\infty}
\frac{ds}{(m_{Z}^{2}-s)s}
+ \int\limits_{s_4}^{\infty}
\frac{\left[
R_{c}^{PT}(s)-R_{c}^{PT}(m_{Z}^{2})
\right] ds}{(m_{Z}^{2}-s)s} \right\} =
\]
\begin{equation}
=  \frac{\alpha^2}{9 \pi}
\left\{ \frac{R_{c}^{PT}(m_{Z}^{2})}{m_{Z}^{2}}
\ln \left( \frac{m_{Z}^{2} - s_4}{s_4} \right)
+ \int\limits_{s_4}^{\infty} \frac{ \left[ R_{c}^{PT}(s) -
R_{c}^{PT}(m_{Z}^{2}) \right] ds}{(m_{Z}^{2} - s) s} \right\}
\end{equation}
%
   For the  $R_{c}^{PT}$ we shall use the formula [10,13]:
\begin{equation}
  R_{c}^{PT}(s) = R_{c}^{(0)} (s) {\cal D} (s)
\end{equation}
where
\begin{equation}
  R_{c}^{(0)} (s) = \frac{3}{2} Q_{c}^{2} v (3 - v^2)
    \; , \; v = \sqrt{1-\frac{4m_{c}^{2}}{s}} \; , \; Q_{c} = \frac{2}{3}
\end{equation}
$m_c = 1.30(5) \; GeV$, [11] and
\begin{equation}
   {\cal D} (s) =
   \frac{4\pi\alpha_s / 3v}{1 - \exp(-4\pi\alpha_s / 3v)} - \frac{1}{3}
   (\frac{\pi}{2} - \frac{3}{4\pi})(3+v) \alpha_s
\end{equation}
In (13) we took into account the terms of the first order in $\alpha_s$
\footnote{Taking into account the terms $\alpha_s^2$ and $\alpha_s^3$
practically does not change $\delta \alpha_{hadr}$}  and "Coulomb"
terms of all orders in $\alpha_s /v$ . At $v \rightarrow 1$
\begin{equation}
    {\cal D} (s) \rightarrow 1 + \frac{\alpha_s (s)}{\pi}
\end{equation}
Note that the function ${\cal D} (s)$ differs from (16) strongly in the region
of resonances (for example ${\cal D} (s_4) = 1.34$ for $J/\psi$ family,
${\cal D} (s_4)= 1.59$ for $\Upsilon$ family), hence it is necessary to
take into account the "Coulomb" term.

For $\alpha_s$ we used formulae B.2-B.4 [14] describing the evolution of
$\alpha_s (s)$ and the new result -
$\alpha_s (m_{Z}^{2}) = 0.125(5)$ [2].

The final result for  $ \delta \alpha_{c}$ can be written in form:

\begin{equation}
 \delta \alpha_{c} =  \delta \alpha_{c}^{Anal.} +  \delta \alpha_{c}^{Num.}
  +  \delta \alpha_{c}^{Reson.}  +  \delta \alpha_{c}^{Add.}
\end{equation}
Where:
\[
 \delta \alpha_{c}^{Anal.} =
   \frac{\alpha}{3 \pi} R_{c}^{PT}(m_{Z}^{2})
   \ln \left( \frac{m_{Z}^{2} - s_4}{s_4} \right) ;
\]

\[
 \delta \alpha_{c}^{Num.} =  \frac{\alpha m^{2}_{Z}}{3 \pi}
    \int\limits_{s_4}^{\infty} \frac{ \left[ R_{c}^{PT}(s) -
   R_{c}^{PT}(m_{Z}^{2}) \right] ds}{(m_{Z}^{2} - s) s} ;
\]

\[
 \delta \alpha_{c}^{Reson.} =  \frac{3 m^{2}_{Z}}{\alpha} \left[
     \sum_{k=0}^{3} \frac{\Gamma_{k}^{ee} M_{k}}{(m_{Z}^{2} - s_k )s_k} +
       \frac{\Gamma_{4}^{ee} M_{4}}{2 (m_{Z}^{2} - s_4)s_4} \right] \;\; ;
\;\; \delta \alpha_{c}^{Add.} =  \frac{1}{8 \alpha} \left(
    \frac{\Gamma_{3}^{ee}}{ M_{3}} - \frac{\Gamma_{5}^{ee}}{ M_{5}}  \right)
\]
Note that numerical calculations show that $\delta \alpha_{c}^{Num.}$ is less
then $\delta \alpha_{c}^{Anal.}$ by a factor of the order $\sim 100$
(see Table 2).
The formulae for the contribution of the $\Upsilon$ - family in $\delta \alpha$
are derived if we replace the index $c$ by $b$ $(m_b = 4.54(2)\; GeV)$
[11].

The main contribution to $\delta \alpha_{hadr}$ comes from the vector
mesons consisting of light quarks. First we consider $\rho$ family.
Instead of (17) we have formulae:
\begin{equation}
 \delta \alpha_{I=1} =  \delta \alpha_{I=1}^{Anal.} +
     \delta \alpha_{I=1}^{Num.}
  +  \delta \alpha_{I=1}^{Reson.}  +  \delta \alpha_{I=1}^{Add.}
\end{equation}
Where:
\[
 \delta \alpha_{I=1}^{Anal.} =
   \frac{\alpha}{2 \pi} \left\{ \left[ 1 + \frac{\alpha_{s}(m_{Z}^{2})}{\pi}
       \right] \ln \left( \frac{m_{Z}^{2}}{s_1} \right) \right\} ;
\]

\[
 \delta \alpha_{I=1}^{Num.} =  \frac{\alpha m^{2}_{Z}}{2 \pi^2}
    \int\limits_{s_1}^{\infty} \frac{ \left[ \alpha_{s}(s) -
   \alpha_{s}(m_{Z}^{2}) \right] ds}{(m_{Z}^{2} - s) s} ;
\]

\[
 \delta \alpha_{I=1}^{Reson.} =  \frac{3}{\alpha} \left(
     \frac{\Gamma_{0}^{ee}}{ M_{0}} +
       \frac{\Gamma_{1}^{ee}}{2 M_{1}} \right) \;\; ; \;\;
 \delta \alpha_{I=1}^{Add.} =  \frac{1}{8 \alpha} \left(
    \frac{\Gamma_{0}^{ee}}{ M_{0}} - \frac{\Gamma_{2}^{ee}}{ M_{2}}  \right)
\]
In eq.(18) the replacement of the summation by an integration has been
carried out starting from $k=1$ , and all terms not written have been
discarded.
Similar equations are valid for $\delta \alpha_{I=0}$ ($\omega$ - family)
and $\delta \alpha_s$ ($\varphi$ - family).
Masses and electronic widths of the $\rho$, $\omega$ and $\varphi$ resonances
are presented in Table 1. Note that in the  $\omega$ case we have two
different sets for the description of the experimental results [15]. The
difference for both cases in value $\delta \alpha_{I=0}$ is negligible.

The results of the calculations are given in Tables 2,3. Summing up the
contributions  of all quarks we get:
\begin{eqnarray}
   \delta \alpha_{hadr} = 0.0275(2) \nonumber \\
    \overline{\alpha} = (128.96(3))^{-1}
\end{eqnarray}
In fact, the uncertainty of $\overline{\alpha}$ is determined
by the largely by the uncertainty of the electronic width
of the $\rho'$ - meson. Therefore more exact measurement of $\rho'$ - meson
electronic width would be very useful.

We thank L.B. Okun and M.I. Vysotsky for useful discussions.

\normalsize{

\section{References}
{\frenchspacing
\begin{tabbing}

1.~~~\=V.A.~Novikov, L.B.~Okun, M.I.~Vysotsky, Mod.Phys. Lett. {\bf A8} \\
\>(1993) 2529, Err. A8 (1993) 3301, \\
\>JETP {\bf 76} (1993) 725, \\
\>Nucl. Phys. {\bf B 397} (1993) 35. \\
\\
2.\>V.A.~Novikov, L.B.~Okun, A.N.~Rozanov, M.I.~Vysotsky, Preprint \\
\>CERN-TH 7217/94 (1994), to be published in Phys. Lett.\\
\\
3.\>B.~Peitzuk, M.~Woods, Talks at 1994 Mariond Conference on "Electroweak\\
\>Interactions and Unified Theories".\\
\\
4.\>V.A.~Novikov, L.B.~Okun, M.I.~Vysotsky, Phys. Lett. {\bf B 324} (1994)
89.\\
\\
5.\>H.~Burkhard, F.~Jegerlehner, G.~Penso and C.~Verzegnassi, Zeit. Phys. \\
\>{\bf C43} (1989) 497. \\
\\
6.\>N.~Cabibbo, R.~Gatto, Phys. Rev. {\bf 124}, (1961) 1577. \\
\\
7.\>F.~Jegerlehner,Villigen preprint PSI-PR-91-08 (1991). \\
\\
8.\>R.B.~Nevzorov, A.V.~Novikov and M.I.~Vysotsky, to be published. \\
\\
9.\>B.V.~Geshkenbein, Yad.Fiz. {\bf 49}, 1138 (1989).\\
\\
10.\>B.V.~Geshkenbein, Yad.Fiz. {\bf 51}, 1121 (1990).\\
\\
11.\>B.V.~Geshkenbein, V.L.~Morgunov, SSCL-Preprint-534, (1993) \\
\\
12.\>Phys. Rev. {\bf 45D} partII (1992) III.54. \\
\\
13.\>Handbook of Mathematical Functions, Edited by M.~Abramowitz and\\
\>I.A.~Stegun, 1964.
\\
14.\>V.A.~Novikov, et al. Phys. Rep. {\bf 41} (1979) 1. \\
\\
15.\>A.B.~Clegg, A.~Donnachie, Preprint M-C-TH-93-21.\\
\\

\end{tabbing}
}
\newpage

\begin{table}[t]

\begin{center}

Table 1. The values of masses and electronic widths of resonances.
\vspace*{0.2cm}

$u,d$ - $\rho$  family [14,15]
\vspace*{0.2cm}

\begin{tabular}{|l|l|l|l|}
\hline
&0&1&2 \\
\hline
$M_{i,Exp.}$ [GeV] & 0.7681(5) & 1.463(25) & 1.73(3) \\
\hline
$\Gamma_{i,Exp.}^{ee}$ [KeV] & 6.77(32) & 2.5(9) & 0.69(15) \\
\hline
\hline
\end{tabular}
\vspace*{0.2cm}

$u,d$ - $\omega$  family [14,15], two variants
\vspace*{0.2cm}

\begin{tabular}{|l|l|l|l|}
\hline
&0&1&2 \\
\hline
$M_{i,Exp.}$ [GeV] &0.78195(14)&1.44(7)&1.606(9) \\
\hline
$\Gamma_{i,Exp.}^{ee}$ [KeV] &0.60(2)& 0.150(38)&0.140(35) \\
\hline
\hline
$M_{i,Exp.}$ [GeV]  &0.78195(14)&1.628(14)& - \\
\hline
$\Gamma_{i,Exp.}^{ee}$ [KeV] &0.60(2)&0.37(10)& -  \\
\hline
\hline
\end{tabular}
\vspace*{0.2cm}

$s$ - $\varphi$  family [14]
\vspace*{0.2cm}

\begin{tabular}{|l|l|l|}
\hline
&0&1 \\
\hline
$M_{i,Exp.}$ [GeV] & 1.019413(8) & 1.70(2) \\
\hline
$\Gamma_{i,Exp.}^{ee}$ [KeV] & 1.37(5) & 0.70(18) \\
\hline
\hline
\end{tabular}
\vspace*{0.2cm}

$c$ - $J$/$\psi$  family [14]
\vspace*{0.2cm}

\begin{tabular}{|l|l|l|l|l|l|l|}
\hline
&0&1&2&3&4&5 \\
\hline
$M_{i,Exp.}$ [GeV]
& 3.09693(9) & 3.6860(1) & 3.7699(25) & 4.04(1) & 4.159(20) & 4.415(6) \\
\hline
$\Gamma_{i,Exp.}^{ee}$ [KeV]
& 5.36(29) & 2.14(21) & 0.26(4) & 0.75(15) & 0.77(23) & 0.47(10)  \\
\hline
\hline
\end{tabular}
\vspace*{0.2cm}

$b$ - $\Upsilon$  family [14]
\vspace*{0.2cm}

\begin{tabular}{|l|l|l|l|l|l|l|}
\hline
&0&1&2&3&4&5 \\
\hline
$M_{i,Exp.}$ [GeV]
& 9.46032(22) & 10.02330(31) & 10.3553(5) & 10.5800(35)  & 10.865(8)  &
11.019(8) \\
\hline $\Gamma_{i,Exp.}^{ee}[KeV] $ & 1.34(4) & 0.56(9) &
0.44(4) & 0.24(5) & 0.31(7) & 0.13(3)  \\
\hline
\hline
\end{tabular}

\vspace*{0.5cm}

Table 2. The contributions to $\delta\alpha(m_{z}^{2})$
\vspace*{0.2cm}

\begin{tabular}{|l|l|l|l|l|l|}
\hline
&  $u,d$ - $\rho$   & $u,d$ - $\omega$ &  $s$ - $\varphi$
&   $c$ - $J$/$\psi$  & $b$ - $\Upsilon$  \\
\hline
 Anal. Int.   &  0.009980(16) & 0.001084(1)  & 0.002137(3)   & 0.006772(10)
 & 0.001165(8) \\
 Resonances   &  0.00397(21)  & 0.000376(15) & 0.000637(29)  & 0.001094(49)
 & 0.000115(5) \\
 Num. Int.    &  0.000016(0)  & 0.000001(0)  & 0.000003(0)   & 0.000030(0)
 & 0.000009(0) \\
 Add. term    &  0.000144(0)  & 0.000000(0)  & 0.000000(0)   & 0.000001(0)
 & 0.000000(0) \\
\hline
              &  0.01412(21)  & 0.00146(2)   & 0.00278(3)    & 0.00790(5)
& 0.00129(1)  \\
\hline
\hline
\end{tabular}

\vspace*{0.5cm}

Table 3. Summarized result
\vspace*{0.2cm}

\begin{tabular}{|l|l|}
\hline
\hline
& $\delta\alpha(m_{Z}^{2})$ \\
\hline
$u,d$ - $\rho$       & 0.01412(21) \\
$u,d$ - $\omega$     & 0.00146(2)  \\
$s$ ~~~- $\varphi$   & 0.00278(3)  \\
$c$ ~~~- $J$/$\psi$  & 0.00790(5)  \\
$b$ ~~~- $\Upsilon$  & 0.00129(1)  \\
\hline
$\delta\alpha(m_{Z}^{2})$        & 0.02754(22) \\
\hline
\hline
\end{tabular}

\end{center}
\end{table}
}
\end{document}